*The Relationship between Image Dynamic Range and Antenna-based Gain Errors*

*ngVLA Memo 107*

T.K. Sridharan, Kumar Golap, Sanjay Bhatnagar & Steve Myers

**Introduction and motivation**

The ngVLA science requirements call for continuum image dynamic ranges of 45 dB and 35 dB at 8 and 27 GHz respectively (SCI0113; SYS6103). In interferometric aperture synthesis imaging, visibility amplitude and phase errors result in errors in the final images, limiting the dynamic range attained. In order to achieve the ngVLA dynamic range requirements it is necessary to limit the amplitude and phase errors to within appropriate levels. In a properly operating instrument, the visibility errors arise from individual antennas – they are antenna based.

Perley (1999, P99 hereafter) derived a nominal relationship between the image dynamic range ideally attained and the visibility phase and amplitude errors causing the image errors. P99 captured essential ideas, providing guidance for high dynamic range imaging and an understanding of the limitations. Hales (2019; H19 hereafter) arrived at the same relationship through the same derivation with a minor modification. In this short memo, we argue that this relationship may not hold everywhere in an image and derive the relationship in a more stringent limit. We modify the P99 and H19 derivations to arrive at the more stringent relationship. Such a relationship is at the root of allowable amplitude and phase errors arising from practically every corrupting effect, e.g. antenna pointing, primary beam characteristics, tropospheric and ionospheric phase fluctuations and polarimetric imperfections, among others (e.g. Braun, 2015). Thus, it is central to ngVLA calibration requirements and strategies.

The essential difference between the P99 relationship and the one derived here is the scaling with the number of antennas in the array, $N$. We derive a more stringent $\sqrt{N}$ dependence as opposed to the previous, generally adopted $N$ scaling and compare with existing simulations. In essence, we conclude that it is more appropriate to consider a dependence over the range $\sqrt{N}$ to $N$. The current ngVLA calibration requirements adopt the less stringent $N$ dependence.

**Current and modified scaling: coherent and incoherent error combination**

We first note general heuristic expectations: a $N$ dependence ($= \sqrt{N^2}$) in the presence of $N^2$ independent baseline-based errors, and similarly, a $\sqrt{N}$ dependence for $N$ independent antennas-based errors, which is the case discussed here.

We only consider antenna-based amplitude and phase errors as the corrupting factors. Baseline based, or equivalently correlator-based, errors are assumed to have been eliminated. We follow and modify the derivations in P99 and H19 to arrive at the $\sqrt{N}$ dependence. A point source with unit visibility on all baselines and a noise free system are assumed, the amplitude and phase errors under consideration being the only errors. The phase and fractional amplitude errors are represented by $\phi$ and $\epsilon$, with $\phi_A, \phi_B, \epsilon_A, \epsilon_B$ being errors per antenna and per baseline. The per-baseline errors relate to small independent per-antenna errors as

$$\phi_B = \sqrt{2}\phi_A \quad (1)$$
$$\epsilon_B = \sqrt{2}\epsilon_A \quad (2)$$

For each baseline, these errors result in a (co)sinusoidal error fringe in the image, of amplitude $\sim \epsilon$ or $\phi$, with the same fringe spacing as applicable to that baseline, being added to the signal fringe of amplitude 1. We are interested in the combination of such error fringes from multiple baselines in the final image.

P99 derives an expression for an $N$ antenna array with $N(N-1)$ ($\sim N^2$) independent baseline-based errors (Eq 13-6 in P99), consistent with the heuristic expectation:

$$DR_{N,Bb} \approx N/\sqrt{2}\phi_B \quad (3)$$

Next, considering error on only one antenna (all other antennas are error free), combining all baselines, the signal is $N^2$, being the sum of unit contributions from $N^2$ baselines. The error is contributed by $N-1$ baselines to the single antenna that suffers the error. In one limit, these antenna-based errors combine incoherently in the image, adding up as $\sqrt{N-1}$. Even though the errors on the $N-1$ baselines arise from a single antenna and therefore are the same on all baselines and are not independent, at different image locations the error fringes may combine incoherently (may not line up) depending on the specific baseline distribution in an array. With a random distribution, $\sim \sqrt{N-1}$ is applicable, leading to Eq 13-7 in P99:

$$DR_{incoh,Ab} = \frac{N^2}{\sqrt{N-1}} \frac{1}{\phi_A} \approx \frac{N\sqrt{N}}{\phi_A} \quad (4)$$

The other limit is obtained with coherent combination, *e.g.* at the location of the source itself. Although image dynamic range at the source location may not have a clear meaning, characterizing fidelity or dynamic range for variability instead, it represents a limiting boundary case. Importantly, this limit is applicable to polarimetric observations at the source location. In this case, the errors would be identical, and therefore, not combine incoherently. The errors add up coherently as $(N-1)$ in the denominator, as opposed to $\sqrt{N-1}$ in the incoherent case. The dynamic range then becomes

$$DR_{coh,Ab} = \frac{N^2}{N-1} \frac{1}{\phi_A} \approx \frac{N}{\phi_A} \quad (5)$$

This step represents the essential deviation from P99.

Moving to the full array from the single antenna case, when all of the $N$ antennas have independent antenna-based errors of similar magnitude of $\sim \phi_A$, the DR scales by $\sqrt{N}$ for the full combination, leading to

$$DR_{N,incoh,Ab} = \frac{N}{\phi_A} \tag{6}$$

for the incoherent case (from Eq 4; same as Eq 13-8 in P99, the final result) and

$$DR_{N,coh,Ab} = \frac{\sqrt{N}}{\phi_A} \tag{7}$$

for the coherent case (from Eq 5), which is the modified result in this memo.

In a minor modification, the effect of independent antenna-based errors from $N-1$ antennas on the $N-1$ baselines to one antenna is considered in H19 (as opposed to error on only one antenna as the starting point in P99). A multiplication factor of $(N-1)/\sqrt{N-1}$ is applied, combining scaling by $N-1$ for the signal, and by $\sqrt{N-1}$ for the error, applicable when the errors add incoherently. Then, combining the whole array, DR improves by $\sqrt{N-1}$, leading to a $N$ dependence (Eq 7 in H19; same result as Eq 8 in P99 and Eq 6 in this memo). However, in the coherent combination case, the errors also add as $N-1$, which makes the multiplication factor above 1 ($= (N-1)/(N-1)$). For the full array, summing over $N$ such antennas with independent antenna-based errors, each with its $(N-1)$ baselines, the DR scales as $\sqrt{N}$, leading again to the more stringent result (same as Eq 7):

$$DR_{N,coh,Ab} \approx \sqrt{N}/\phi_A \tag{8}$$

The dynamic range for amplitude errors may be obtained by replacing $\phi_A$ with $\epsilon_A$.

To summarize, the image dynamic range scales as $\sqrt{N}$ or $N$ depending on coherent or incoherent combination of the effects of the visibility errors in the image. Coherent error combination leads to a lower dynamic range and equivalently, more stringent antenna-based amplitude and phase error limits to achieve a given dynamic range. Conversely, for a given visibility error, the dynamic range achieved would be lower by 10-12 dB for 107 - 214 antennas, compared to the levels with $N$ dependence currently adopted in the ngVLA calibration requirements. To cover all situations, it is necessary to consider a $\sqrt{N}$ to $N$ scaling range and only a conservative $\sqrt{N}$ for deriving the requirements.

**Comparison with simulations**

Hales (2020; H20 hereafter) conducted simulations focused on amplitude errors arising from pointing errors to characterize the achieved dynamic ranges and to verify the $N$ dependence. While H20 concluded that the simulation results supported a $N$ dependence, we show here that the results are more consistent with a $\sqrt{N}$ dependence, in particular for the levels of pointing errors expected to be applicable to the ngVLA antennas (3″ under precision conditions).

H20 compares simulation results with expectation from simple theory (fig 5 in H20), adapted and reproduced as fig 1 here. The H20 comparison is compromised by (1) while the simulations are for 214 antennas, the best matching theory shown only considers 107 antennas as a way to account for the use of natural weighting in the simulations. However, since the theoretical derivations treat all baselines equally which corresponds to natural weighting, all 214 antennas should be included in comparing with the natural weighted simulation results (2) the amplitude error incorporated in the theory uses the power beam of the antenna (Eq 2 in H20), rather than the applicable voltage beam for antenna based errors. Fig 1 accounts for the above shortcomings. (3) the computational dynamic range limit of the simulations is ~ 42 dB compared to the 45 dB ngVLA requirement at 8 GHz. However, the simulations were carried out at 27 GHz and meet the corresponding lower 35 dB ngVLA requirement. The results should still be applicable to 8 GHz, with the pointing error in the x-axis scaled for frequency and represented as a fraction of the beam size, rather than arcseconds.

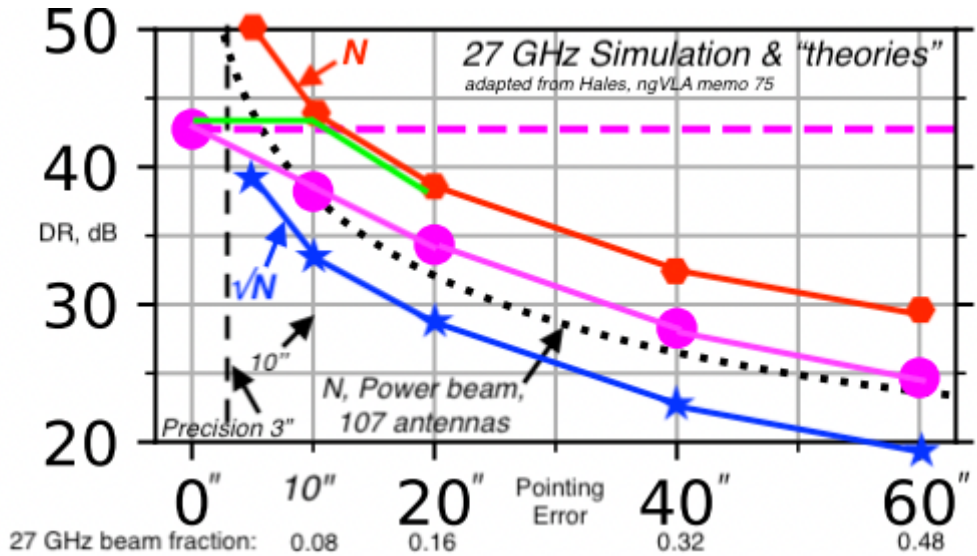

*Figure 1. Comparison of the expectation from simple theory with results of simulations for image dynamic range (adapted from Fig 5 of Hales 2020, H20). Image dynamic ranges achieved (DR; y-axis) with different pointing errors (x-axis) for an on-axis point source are depicted. The simulation results at 27 GHz for 214 antenna are shown by megenta circles. The black dashed curve shows theory as plotted in H20, for 107 antennas, power beam and N dependence. The red hexgons show N dependence and (corrected) theory for 214 antennas and voltage beam, as correctly applicable to the simulations. They provide a much poorer macthing to the simulations than the black dashed curve. The blue pentagrams show corrected theory (i.e. voltage beam, 214 antennas) and the $\sqrt{N}$ dependence derived in this memo. The current ngVLA 18-m pointing error specifications under precision and normal conditions are marked (3″ & 10″; ~ 0.1-0.3 × primary beam at 116 GHz). The green line shows the expected saturation profile for N dependence due to the dynamic range limit of the simulations, which is not seen in the simulation results (magenta circles). The simulation results fall between N and $\sqrt{N}$ dependence. In the relevant ≲ 10″ pointing error range, applicable to ngVLA, the $\sqrt{N}$ dependence provides a much better match to the simulation results. A comparison of the behaviour of the theory and simulation outside the ≲ 10″ region is not very meaningful. Additionally, note that in linear scale the $\sqrt{N}$ curve would be closer to the simulation results than the N curve everywhere.*

Comparison with simple theory for validation is appropriate for the simplest boundary case of an on-axis point source and small pointing errors, where the theory is applicable, as presented in Fig 1. As can be seen, the simulation results fall in the $\sqrt{N}$ to $N$ range. In the relevant small pointing error limit ($3'' - 10''$; $\sim 0.1 - 0.3 \times Primary\ Beam\ FWHM$ at 116 GHz ), a $\sqrt{N}$ dependence provides a better match. Considering the dynamic range limit of the simulations, a dynamic range saturation is expected for $N$ dependence. Indicated by the green line for the small pointing error region (not a rigorous calculation), this saturation is clearly not seen to match the simulation results, particularly in comparison to the $\sqrt{N}$ dependence (blue) where such a saturation is not expected (due to the theoretical dynamic range not approaching the dynamic range limit of the simulations).

**Conclusion**

We conclude that in order to achieve the dynamic range requirements in a range of observations, covering regions at or near the phase center (or the location of the brightest sources in the field) and over the full primary beam, the ngVLA must target amplitude and phase error limits implied by the more stringent $\sqrt{N}$ dependence. Adopting a $\sqrt{N}$ dependence is adequate at this time, based on the derivation and arguments presented above. In the longer term, expanded simulations incorporating expected aperture illumination, a variety of errors, a higher simulation dynamic range limit and better sampling of the $< 20''$ region, which we intend to conduct, would be useful.